\documentclass[preprint,12pt]{aastex}

\slugcomment{Submitted to {\it{The Astrophysical Journal}}}
\shortauthors{Bernstein, Freedman, Madore}
\shorttitle{Optical EBL: Addendum}

\newcommand \gs{\mathrel{\raise0.35ex\hbox{$\scriptstyle >$}\kern-0.6em
                \lower0.40ex\hbox{{$\scriptstyle \sim$}}}}
\newcommand \ls{\mathrel{\raise0.35ex\hbox{$\scriptstyle <$}\kern-0.6em
                \lower0.40ex\hbox{{$\scriptstyle \sim$}}}}

\begin{document}

\title{ 
Corrections of Errors in `The First Detections of the Extragalactic
Background Light at 3000, 5500, and 8000 \AA. I, II, and III'
}

\author{
Rebecca A.\ Bernstein,\altaffilmark{1} 
Wendy L.\ Freedman,\altaffilmark{2}
and Barry F.\ Madore\altaffilmark{2}}

\altaffiltext{1}{University of Michigan, Ann Arbor, MI, 48109}
\altaffiltext{2}{Carnegie Observatories, 813 Santa Barbara St, Pasadena, CA, 91101}

\setcounter{footnote}{3}

\begin{abstract}
We correct errors in Bernstein, Freedman \& Madore (2002abc), a series
of papers in which we described observations of the optical
extragalactic background light (EBL). These errors pertain to the
measurement of zodiacal light, given in the second paper of this
series.  Making these corrections leads to a net decrease of
0.5($\pm$0.6)\% in our zodiacal light measurement and a corresponding
increase in the inferred extragalactic background light of roughly
$0.5 (\pm 0.6) \times 10^{-9}$ ergs sec$^{-1}$ cm$^{-2}$ sr$^{-1}$
\AA$^{-1}$.  For comparison, the originally--quoted EBL flux at
5500\AA\ was $2.7 (\pm1.4) \times 10^{-9}$ in the same units
($1\sigma$ combined systematic and statistical uncertainty).  We
provide a detailed discussion of these errors and also discuss the
evolution of this work prior to the (2002) papers.  We note that
corrections of the factual errors in our (2002) papers yield a result
that is consistent with the results and errors quoted there.  However,
this is not intended to be a new or updated analysis, and it does not
address some methodological objections which have been raised to our
prior work.
\end{abstract}

\keywords{general,galaxies:background light}

\section{Introduction}

In Bernstein, Freedman \& Madore (2002a,b,c; hereafter, BFM1, BFM2 and
BFM3, respectively), we described a measurement of the mean flux of
the extragalactic background light (EBL) in a $5$ arcmin$^2$ field of
view.  In that study, the EBL contribution was identified by measuring
the total flux in the target field and subtracting from it the flux
contributed by known foreground sources, namely diffuse Galactic light
(DGL) and zodiacal light (ZL):
\begin{equation}
I_{\rm EBL} = I_{\rm tot} - I_{\rm ZL} -  I_{\rm DGL} . 
\end{equation}
The total flux, $I_{\rm tot}$, was measured from space using
HST/WFPC2 imaging in the $U-$, $V-$, and $I-$bands and using HST/FOS
spectroscopy covering roughly 4000--7000\AA. The zodiacal light,
$I_{\rm ZL}$, was measured using ground--based spectrophotometry
obtained at the du Pont 2.5~m telescope at Las Campanas Observatory
(LCO) in Chile.  The diffuse Galactic light, $I_{\rm DGL}$, was
estimated from a simple scattering model. The EBL values at 3000,
5500, and 8000\AA ~were measured to be 4.0 ($\pm$2.5), 2.7 ($\pm$1.4),
and 2.2 ($\pm$1.0) $\times 10^{-9}$ ergs sec$^{-1}$ cm$^{-2}$ sr$^{-1}$
\AA$^{-1}$, respectively. 

The zodiacal light measurement, described in BFM2, is the only part of
the experiment involving ground--based observations.  The Earth's
atmosphere influences these observations through absorption,
scattering, and airglow emission.  Absorption and scattering cause
``extinction'' of the light in the target field; scattering causes a
fraction of the light from the full, visible hemisphere of the sky to
be added to the line of sight; and airglow is an additive foreground
source produced in the atmosphere.  The resulting
night--sky spectrum observed from the ground ($I_{\rm NS}$) was
therefore described as a function of time($t$), airmass ($\chi$),
atmospheric extinction ($\tau_{\rm obs}$), and wavelength ($\lambda$),
as follows (Equation 3 of BFM2):
\begin{equation}
I_{\rm NS}(\lambda,t,\chi) = 
I_{\rm target} e^{-\tau_{\rm obs}(\lambda)\chi}  +
I_{\rm scat}(\lambda,t,\chi) + 
I_{\rm air}(\lambda,t,\chi),
\end{equation}
in which $I_{\rm scat}$ is the spectrum of light scattered into the
line of sight and $I_{\rm air}$ is the airglow spectrum.  To measure
the zodiacal light in our experiment, we utilized the fact that the
zodiacal light is known to have a slightly--reddened Solar spectrum in
which the Solar--strength Fraunhofer lines are preserved. The zodiacal
light contamination can therefore be expressed as the product of a
fiducial Solar spectrum, $I_\odot(\lambda)$, and a scaling function,
$C(\lambda)$, that varies roughly linearly with wavelength.  The
airglow spectrum, on the other hand, is not known to contain
Fraunhofer features.  We therefore identified the ZL flux by
iteratively determining the scaling factor, $C(\lambda)$, for which
the resulting residual airglow spectrum has the minimum correlation
with the Solar spectrum. We expressed the airglow spectrum as
\begin{equation}
I_{\rm air}(\lambda,t,\chi) = 
	 I_{\rm NS}(\lambda,t,\chi) - I_{\rm ZL}\left[ 
	     e^{-\tau_{\rm obs}(\lambda)\chi} 
	     + \left( 
	\frac{I_{\rm EBL}(\lambda)+ I_{\rm DGL}(\lambda)}{I_{\rm ZL}}
	\right)  
	 e^{-\tau_{\rm obs}(\lambda)\chi}
	  + \frac{I_{\rm scat}(\lambda, t,\chi)}{I_{\rm ZL}}
	   \right],
\end{equation}
in which the term $e^{-\tau_{\rm obs}(\lambda)\chi}$ accounts for ZL
flux lost from the beam due to extinction and the scattered light
term, $I_{\rm scat}$, includes ZL, ISL (integrated star light), and
DGL (diffuse Galactic light) as contaminating sources.  As discussed
in BFM2, we then needed a model for each scattering source over the
visible spectrum in order to calculate the scattered light. To
eliminate the absolute flux of the ZL from the models, we expressed
the scattered light from all sources as a fraction of the ZL in the
target field, as implied by Equation 3. We then combined the net ZL
loss due to extinction with net ZL gain due to scattering to give an
effective extinction, $\tau_{\rm eff}(\lambda)$. This let us express
the effect of the atmosphere on the ZL as a relative (multiplicative)
correction.  The absolute value of the ISL, DGL, and EBL remain in the
calculation.  However, the EBL and DGL terms were then dropped because
they were not expected to have Fraunhofer features and so were not
expected to impact the spectral measurement based on the strength of
these features.\footnote{This is correct for the EBL, however the DGL
  can and does contribute Fraunhofer features and its spectrum should
  be included as a contribution to the target field and as a source of
  scattering in the spectral measurement of the ZL.  The strength of
  zodiacal Fraunhofer lines in the DGL 
  is weaker than in the solar spectrum by a factor of three, so that
  the impact on our measurement would be roughly $0.01\times I_{\rm
  ZL}$.}
We therefore identified the ZL flux (expressed as $C(\lambda)I_\odot$)
according to the equation:
\begin{equation}
I_{\rm air}(\lambda,t,\chi) =
I_{\rm NS}(\lambda,t,\chi) - C(\lambda)I_\odot   
\left[ 	 e^{-\tau_{\rm eff}(\lambda)\chi}  
    + \frac{I^{\rm ISL}_{\rm scat}(\lambda, t,\chi)}{I_{\rm ZL}}
    \right]. 
\end{equation}

We correct errors in our (2002) papers regarding the dates of the Las
Campanas Observatory (LCO) observations, a statement regarding the
location of the Moon on those dates, and quantify the implications of
these corrections. We also include a discussion of analysis errors
which pertains to all unrefereed work prior to the (2002) papers
(Bernstein, Freedman, \& Madore 1996; Bernstein \& Madore 1997;
Bernstein 1998PhDT; Bernstein 1999a; Bernstein 1999b); these were
corrected before publication of BFM2 and lead to no corrections here.
Some of these errors were noted by Mattila (2003).  We adopt
nomenclature consistent with that of our earlier work to allow clear
discussion of what was done there. Throughout this paper, we
abbreviate $10^{-9}$ ergs sec$^{-1}$ cm$^{-2}$ sr$^{-1}$ \AA$^{-1}$ as
cgs.

\section{Errors in Early Analysis}

Mattila (2003) has correctly noted that the analysis of the
ground--based data as detailed in the unpublished thesis (Bernstein
1998) contained an incorrect treatment of atmospheric effects.  In
that early analysis, atmospheric scattering was not included as a
contribution to the observed night sky spectrum.  In addition, due to
a programming error in a subroutine, an incorrect extinction
correction was applied.  The incorrect treatment of atmospheric
scattering was identified by the referee and both errors were
corrected before publication.  In the unpublished thesis, prior to the
correction of these two errors, the ZL was therefore calculated based
on the following expression (compare to Equation 4 above):
\begin{equation}
I_{\rm air}(\lambda,t,\chi) =
I_{\rm NS}(\lambda,t,\chi) - 
 C(\lambda)I_\odot 
e^{-[\tau_{\rm obs}(\lambda)- \tau_{\rm obs}(4600)]\chi}.
\end{equation}
Over the wavelength range used in that analysis (4200--5100\AA) and at
the mean airmass of our observations ($\chi\sim$ 1.2),
the exponential term in Equation 5 has values between 0.93 and 1.06,
with an average value of 1.00.  In effect, the data were analyzed with
no scattering correction and an incorrect extinction correction.

In brief, the analysis in Bernstein (1998) involved preparing solar
spectra appropriate to each observation using IRAF routines to
resample in wavelength, and apply an extinction curve for the airmass
of each observation. The zodiacal light solution is then a scaling
value relative to these prepared fiducial solar spectra.  The solar
spectra, corrected for extinction, were compared with the
corresponding LCO spectra to find the contribution of zodiacal
light. These solar spectra were produced many times in the course of
data reduction, because the extinction and sensitivity solutions were
recalculated many times. A check was therefore included in a subroutine
to confirm that the solar spectra were correctly prepared.  That check
involved multiplying the solar spectrum by $\exp[\tau_{\rm
obs}(4600)\chi]$ with $\tau_{\rm obs} (4600)= 0.2$ mag/airmass, roughly
removing the extinction correction. The error then occurred by passing
the wrong vector back to the main program from the subroutine.  The
programming error was not identified until the anonymous referee
pointed out the incorrect treatment of atmospheric scattering. For
completeness, we note that Figure 4.4 in Bernstein (1998) does not
show the final extinction solution used in the thesis.  The correct
extinction solution used in all versions is shown in BFM2. Table 2.9
in Bernstein (1998), which lists values for $I_{\rm tot}$, was also
updated in the published papers.

When properly treated, the scattering and extinction are nearly equal
in magnitude but opposite in sign, and so they cancel to a high degree
(to about 0.5\% averaged over wavelength and airmass), giving the same
result as the original analysis to within the accuracy of the
measurements.  The cancellation of the scattering and extinction terms
in the proper analysis can be seen in the following quantitative
example. At 4600\AA\ and at our mean airmass (as given above), the
extinction coefficient (Figure 29 of BFM2) is $\tau_{\rm
eff}(4600)=0.042$~mag/airmass and, accordingly, $\exp[-\tau_{\rm
eff}(\lambda)\chi]$ = 0.955.  Over the spectral range 3900--5100\AA\
and at the same airmass, the net flux gained from scattered ISL is in
the range 10--17 cgs.  At 4600\AA, the ISL scattered flux is 12.5 cgs.
The ISL contribution impacts our ZL measurement to the degree that it
contributes to the strength of the Fraunhofer lines in the observed
night sky spectrum; however, those features are only 10\% to 40\% as
strong in the scattered ISL as in the ZL over the range 3900--5100\AA,
and 30\% to 40\% as strong around 4600\AA\ (see Figures 29 and 30 of
BFM2). The scattered ISL therefore contributes +4.4 cgs (=12.5 cgs
$\times$ 0.35) to the solution, or $0.040\times I_{\rm ZL}$ (given
that $I_{\rm ZL}$ is roughly 110 cgs.)  The term in square brackets in
Equation 4 is therefore nearly unity (0.995 for this example).  At
higher airmasses and shorter wavelengths, scattered flux ($I_{\rm
scat}$) and extinction ($\tau_{\rm eff}$) both increase. At smaller
airmasses and longer wavelengths, they both decrease.  In either case,
the term in square brackets in Equation 4 is still nearly unity.  In
other words, the net loss due to extinction and the net gain due to
scattering are synchronized and cancel to a level that is much smaller
than the uncertainty in identifying the ZL flux contribution in the 16
spectra used in this analysis, which have an {\it rms} scatter of
2.3\%.  Because of this cancellation, statistically indistinguishable
results were obtained in the early (Bernstein 1998) and published
(BFM2) versions of the analysis.  Note that the similarity between the
net effects of atmospheric scattering and extinction alone are
coincidental, and would likely not occur along lines of sight where
the ZL in the target field is much brighter or fainter. They are also,
of course, dependent on the parameters used to calculate the
scattering model, which are documented explicitly in BFM2.

In the early analysis, no change in $C(\lambda)$ with wavelength was
detected because the incorrect extinction correction masked the
reddening of the ZL relative to the Sun.  Because no color term was
detected, several broad bandpasses were used.  In the published
version, an increase in $C(\lambda)$ with wavelength was identified,
consistent with the reddened color of the ZL relative to the solar
spectrum.  Narrow bandpasses focused on the solar features were then
used to help identify this trend.

\section{Dates of Ground--Based Observations.}

The dates of the ground--based observations were incorrectly recorded
in the unpublished thesis and were subsequently transcribed by RAB
from there into the published papers. The original observing logs and
the records of the observatory show that the correct dates of the run
were the local--time nights of 23/24 November 1995 through 27/28
November 1995 (5 nights total).  The last night of the run was used
for imaging.  Data from the first and third nights were not used due
to weather and mechanical problems, as described in Bernstein (1998)
and BFM2.  The spectra cited and analyzed in BFM2 were therefore taken
on nights 2 and 4 of the run, having local--time dates 24/25 and 26/27
November 1995.  The corresponding UT dates were 25 and 27 November
1995.  The incorrect dates affect the application of the zodiacal
light measurement to the HST observations at the level of 0.2\%
(although with significant uncertainty) and also affect the scattering
calculations at the level of $<0.1$\%.  We describe and quantify these
two effects below.

\subsection{Relevance for HST Observations}

The HST observations analyzed in BFM1 were executed on the UT nights
of 29 November 1995 and 16--17 December 1995.  Ground--based
observations were assigned and scheduled by the time allocation
committee about one year earlier.  As stated in the abstract of BFM1,
the observations were designed to occur contemporaneously with one of
the sets of HST observations, but they were not executed
simultaneously.

To get an idea what the change in the ZL value might be between our
LCO observations on 25/27 November 1995 and the HST observations on 29
November 1995, we can look at data in the literature and our own HST
data.  From 29 November 1995 to 16/17 December 1995, the HST/WFPC2 and
FOS data {\it both} showed a $2$\% decrease in the mean surface
brightness of the EBL target field.  As discussed in BFM1 and BFM2,
this difference is what would be expected in sign and magnitude as the
heliocentric longitude ($\lambda-\lambda_\odot$) of the target field
goes from about
$150^{\circ}$ on 29 November 1995 to $130^\circ$ on 16/17 December 1995.
One expects this small decrease in intensity because the ZL is
slightly brighter in the anti-solar direction ($\lambda-\lambda_\odot
= 180^\circ$) and has a broad minimum at $\lambda-\lambda_\odot =
130^\circ$.

For comparison, several data sets are available in the literature.
The only all--sky measurements of the ZL surface brightness are from
the ground.  Of these, the most reliable are those tabulated by
Levasseur--Regourd \& Dumont (1980, hereafter LRD80) from their
1964--1975 observations at Tenerife Observatory.  That data set is
reproduced in Leinert et al.\ (1998), where it is updated with
space--based values within 30 degrees of the sun. Although these data
are ground--based and subject to scattering corrections, they are in
good agreement with space--based results, as discussed in LRD80,
Leinert et al.\ (1998), and BFM2.  Between $\lambda-\lambda_\odot
=150^\circ$ and $130^\circ$ and ecliptic latitude $31^\circ-35^\circ$,
the data tabulated in LRD80 show a 
--6\% change in the ZL flux.  At these latitudes, data are also
available from several other sources. As compared and discussed in
Leinert et al.\ (1981) and Leinert et al.\ 1998), Frey et al.\ (1970)
find a change of about 
$+2$\% over these same angles, and the Helios space probes (Leinert et
al.\ 1981, Leinert et al.\ 1982) find a change of 
--1\%.  These three published results are in good agreement to within
the errors of any of the measurements, which are of order 5--10\%.

To be conservative in estimating the change in the ZL between 29
November 1995 and 16/17 December 1995, we simply average the values
discussed above (--2\%, --6\%, $+2$\%, --1\%) to obtain --1.7\% with a
standard deviation of 3.3\%.  As the errors are probably systematic,
the standard deviation may be more indicative of the uncertainty than
the error in the mean.  We then estimate that the change in the
zodiacal light between 25/27 November LCO observations
($\lambda-\lambda_\odot$ = 153 and 151) and 29 November for Hubble
Space Telescope (HST) observations ($\lambda-\lambda_\odot$ = 149)
should be $-0.2 (\pm0.3)$\% (i.e., slightly {\it fainter} on the
29th).  To conservatively allow for any systematic uncertainties
between the data sets, we double this error bar to $\pm0.6$\%. We
include this offset in the summary in Table 2.

\subsection{Relevance for Scattering Calculations}

Because of the transcription error in the dates of the observations,
the scattering calculations in BFM2 were performed for the wrong date,
namely the 29th rather than the 25th and 27th of November 1995.  The
sky visible at a particular UT time shifts by roughly one degree per
day. However, the target field and all sources of scattering obviously
move in consort, so that the scattering calculated for a given zenith
angle of the target is correct on any date. 
The one source which does not move in consort is the moon; however,
any spectrum affected by moonlight should not be included in the
analysis (see \S\ref{moon}), and so the moon is not included in the
calculation of scattered light from extraterrestrial sources.  
The only change in the scattering calculations between November 25,
27, and 29 is therefore caused by the fact that the target field will
rise 4 minutes earlier on each successive night. The scattering
calculations for UT=2:00 on November 29 are therefore correct for
UT=2:16 on November 25, and UT=2:08 on November 23.  The net change in
the scattered ZL and ISL for the largest difference in timing (16
minutes) is very small. Moreover, as illustrated in \S2, the net
effect of the atmosphere (scattering and extinction of ZL, and
scattering of ISL) nearly cancels at every zenith angle.  For that
reason, the change over 8 or 16 minutes is not detectable.

To illustrate this quantitatively, we note that the mean change in the
effective extinction (Figure 25, BFM2) is smaller by an average of
0.0013 mag/airmass between a given UT time and 16 minutes earlier.
This translates into a fractional change of 0.03 in $\tau_{\rm eff}$.
Using the mean airmass, $\chi=1.2$, this corresponds to an increase in
the net ZL by a factor of 1.0015.  As the scattered ISL gets brighter
with increasing UT time, the scattered ISL would be correspondingly
fainter by about 4\% over that same time period (16 minutes earlier),
and would decrease the strength of the ISL spectral features by the
factor 0.998 (= $0.04\times 0.35\times$ 12.5 cgs/$I_{\rm ZL}$).  The
net change with time is then 0.998$\times$1.0015 $=$ 0.9995, which is
not significant.  Nevertheless, for completeness, we list this term,
and all other corrections discussed here in the summary in Table 2.

\section{Location of the Moon}
\label{moon}

\begin{figure}[t]
\plotone{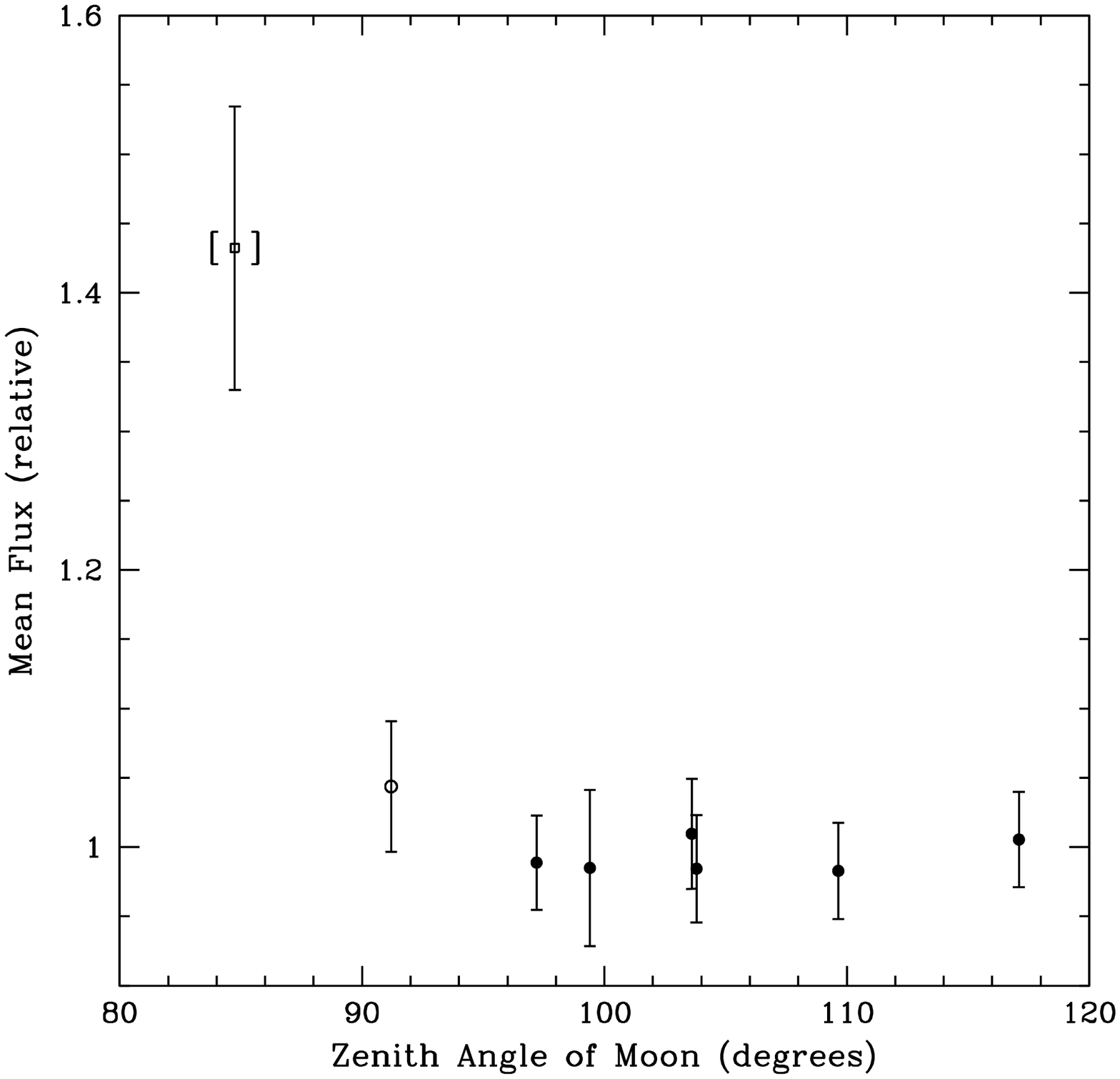}
\caption{\small The mean flux of each exposure listed in Table 1 as a
     function of the average lunar position during the exposure.
     Circles indicate the mean flux of the seven spectra included in
     the analysis in BFM2, normalized by their combined mean. The open
     circle corresponds to the spectrum which was probably 
     affected by moonlight and which we are now excluding from the
     analysis. The point marked by the open square and enclosed in
     brackets indicates the mean flux of the one exposure listed in
     Table 1 which was {\it not included} in the analysis of BFM2;
     this point is normalized by the same value as the other seven.
     To indicate the variability of the spectra as a function of
     wavelength, the error bars show the standard deviation of the
     difference between each spectrum and the mean of the seven
     analyzed spectra.  These differences are due to changes in the
     relative contributions of airglow, scattered light, and ZL from
     the target field.}
\end{figure}

\begin{table}[t]
{\scriptsize
\centerline{\sc Table 1. -- Location of the Crescent Moon During Individual Exposures}
\begin{center}
\begin{tabular}{cccccc}
\hline\hline
\noalign{\smallskip}
Date & Start Time & Exposure Time & Target Field & Lunar Position          & Predicted Moonlight \cr
UT   & UT	  & sec           & Zenith Angle (deg)& Zenith Angle(deg)  & (\% ZL) \cr
\hline
\smallskip
11/25/95	 & 2:18	 & 900	 & $18.3-15.4$  &  $97.8\ $ to\ $100.0$	 &  $<1$\%     \cr
11/25/95	 & 2:34	 & 1800	 & $15.2-10.4$  & $100.9\ $ to\ $106.3$	 & ...    \cr
11/25/95	 & 3:54	 & 1800	 & $10.5-15.4$  & $114.8\ $ to\ $119.4$	 & ...   \cr
&&&&&\cr
\smallskip
(11/27/95)	 &(2:37) &(1800) & $(13.3-9.4)$ & $(81.7\  $ to\ $87.8)$ & ($\sim60$\%)\cr
11/27/95	 & 3:10	 & 1800	 & $9.1-9.8$    &  $88.4\  $ to\ $94.0$	 & $\sim11$\% \cr
11/27/95	 & 3:41	 & 1800	 & $10.0-14.4$  &  $94.7\  $ to\ $99.7$	 & $<1$\%     \cr
11/27/95	 & 4:14	 & 1800  & $15.0-20.9$  &  $101.3\ $ to\ $106.3$ & ...       \cr
11/27/95	 & 4:48	 & 1800	 & $21.7-28.1$  & $106.9\  $ to\ $112.4$ & ...       \cr
&&&&&\cr
\hline
\end{tabular}
\end{center}
}
\end{table}

The LCO data were obtained by RAB and took place several days after
new moon. Each night, as the Moon was setting, the open--dome time was
used to observe bright, standard stars for calibration.  Observations
of the EBL target field began as the Moon approached the horizon.  The
general, but quantitative, statement in BFM2 that the Moon was at
least $14$ degrees below the horizon during all but one exposure is
incorrect.  The correct statement is that one exposure that was used
in the analysis began as the Moon was still setting (UT = 03:10 on 27
November 1995); all other exposures used in the analysis were taken
with the Moon below the horizon by several degrees.  The times of the
first few exposures and the corresponding position of the Moon during
those exposures are given in Table 1 for all spectra taken until the
Moon was more than 22 degrees below the horizon on November 25 and 27.
The spectra from these exposures are plotted in Figure 1.

Because of its high mean flux level, it was clear to us that the
exposure beginning at UT = 02:37 (open square in Figure 1) was
affected by moonlight and for that reason it was not used in the
analysis in BFM2.  The moon contributes exponentially less light with
time after passing below the horizon (like the sun at sunset). The
remaining exposures were not obviously affected and were therefore
included in the subsequent analysis. We now consider what the impact
of the Moon might have been on the included spectra.

To obtain a theoretical estimate of the scattered moonlight which
might influence each exposure, we can use the scattering model
described in BFM2.  These estimates are given in Table 1 as a fraction
of the ZL flux in the target field.  Simpler models for moonlight sky
brightness, such as that implemented by Skycalc.V5 (Thorstensen 2001),
give consistent values at zenith angles smaller than about $85^\circ$,
but yield higher values for the sky brightness very near the
horizon. (These models do not predict the moonlight below the
horizon.  See Krisciunas \& Schaefer (1991) for a discussion of the
model implemented in Skycalc.V5 and its uncertainties.)  For the lunar
phase and angular distance of the target from the Moon ($\sim90^\circ$
on 27 Nov. 1995), the estimated moonlight flux is negligible ($<<1$\%)
by the time the Moon reaches a zenith angle of $98^\circ$.

We can also obtain an empirical estimate of the scattered moonlight in
each exposure by simply comparing their mean fluxes. For the exposure
beginning at UT = 02:37 (which was clearly affected by moonlight and
was {\it not} used in the analysis), the scattered moonlight is
estimated to be about $0.60 \times I_{\rm ZL}$ at 4600\AA.  The ZL
accounts for roughly two thirds of the night sky flux (see Figure 9,
BFM2), so that this spectrum is predicted to be about 40\% brighter
due to moonlight than spectra taken later.  This is generally
consistent with the empirical mean flux of the spectrum, which appears
to be about 43\% brighter than later spectra (open square, Figure 1).
For the exposure beginning at UT = 03:10 (27 Nov. 1995), the scattered
moonlight is estimated to be $0.11\times I_{\rm ZL}$, implying that
the mean flux for this spectrum should be about 8\% higher than later
spectra.  The flux of this spectrum appears to be about 5\% higher
than the mean (open circle, Figure 1), which is again generally
consistent to within the accuracy of the scattering models at very
high airmasses.  Note also that there is about 5\% peak--to--peak
variation in the mean flux of spectra that are not influenced by
moonlight. This is presumably due to changes in atmospheric effects
(changes in airglow, changes in extinction with airmass, and changes
in scattered starlight and diffuse galactic light).  For this reason,
the spectrum at UT=03:10 did not obviously appear to be problematic.

We conclude from the predicted and empirical fluxes of the spectra
discussed above that the spectrum taken at UT=03:10 was probably
affected by moonlight, and so we recalculate the final result without
it.  The ZL value derived from that exposure alone is $113\pm 3$
(1$\sigma$), which is about 3--4\% higher than the mean (see Figures
12 \& 13, BFM2).  Excluding this data point, the final ZL result
(based on the average of 16 observations) is lower by 0.3\%, which is
roughly 1/2 the quoted statistical uncertainty and 1/4 the systematic
uncertainty.  The effect of excluding this exposure from the analysis
is included in the summary in Table 2.

\section{Summary}

We have presented corrections to the published results (BFM1,2,3),
including the dates of the ground--based observations, the location of
the Moon during each exposure, and the quantitative impact of these
corrections.  In addition, we have explicitly documented corrections
made to the analysis between the unpublished thesis (Bernstein 1998)
and published versions of this work (BFM2).  The measured value of the
ZL decreases by 0.5($\pm$0.6)\%, or 0.5($\pm$0.6 cgs) as a result of
these changes. For comparison, the quoted measurement uncertainties in
BFM2 are 0.6\%, statistical, and 1.1\%, systematic.  The inferred EBL
increases correspondingly by $0.1$, $0.5$, and $0.7$ cgs in the $U$,
$V$, and $I-$bands. For comparison, the quoted $1\sigma$ uncertainties
in each band were 2.5, 1.4, and 1.0 cgs, respectively.  The
corrections discussed here therefore yield a result that is consistent
with the previously quoted result and errors; however, this is not
intended to be a new or updated analysis.

\acknowledgments
We thank K. Mattila for his work and comments regarding these results.

\begin{table*}[t]
{\scriptsize
\centerline{\sc Table 2. -- Corrections to Calculated Zodiacal Light}
\begin{center}
\begin{tabular}{l c c c c c} 
\hline\hline
\noalign{\smallskip}
Issues for & BFM2  & Current & Multiplicative  	 & Term(s) 	 & $\Delta I_{\rm ZL}$	\cr 
ZL result  &       &  Version  & Change		 & Affected      & ($\times I_{\rm ZL}$) \cr 
\hline
\smallskip
Shift in ZL (Nov 25,27 -- Nov 29) (\S3.1)
	 & ...
	 & $- 0.2$ cgs
	 & ...
	 & ...
	 & $- 0.002 (\pm 0.006)$  \cr
Date used in Scattering Calculation (\S3.2)
	 & 11/29 & 11/25, 11/27   
	 &  0.966, 0.96 		
	 & $\tau_{\rm eff}$, $I_{\rm scat}({\rm ISL})$
	 & $<0.001$ \cr
Exposure 11/27/95 UT=3:10 (\S3.3)
	 & included 
	 & excluded  	
	 &  $...$ 		
	 &  $...$
	 & $- 0.003$  \cr
\hline
Cumulative Change
	 &	
	 &
	 &
	 &
	 & $-0.005 (\pm 0.006)$  \cr
\hline
\end{tabular}
\end{center}
}
\end{table*}


\end{document}